# A SUBSAMPLE TIME-DELAY ESTIMATOR WITH A FARROW STRUCTURE

*Hugh L. Kennedy*
Adelaide University

## ABSTRACT

A novel solution to the classical problem of subsample time-delay estimation (TDE) for a wideband signal received by two adjacent sensors is described. A mean squared-error (MSE) measure of dissimilarity is defined and evaluated using a variable fractional-delay (VFD) filter bank with a finite impulse response (FIR). Using a filter bank with a Farrow structure allows the MSE power to be expressed as a polynomial function of the filter delay with a local minimum at the unknown signal delay. In Monte-Carlo simulations, the proposed estimator outperforms the generalized cross-correlation (GCC) technique when the signal-to-noise ratio (SNR) is not low (less than 10 dB).

***Index Terms***— Time-delay estimation, variable fractional-delay filters, Farrow structure, cross correlation

## 1. INTRODUCTION

This paper presents a new technique for time-delay estimation (TDE) involving a wideband signal received by a pair of adjacent sensors. The sensors are sampled uniformly, sequentially, and simultaneously, but the sampling period is greater than the time-delay due to the speed of signal propagation, thus subsample estimates are required. Such cases arise when the data-acquisition rate is limited by the data-processing rate (due to weight and power constraints) and/or when the sensors are closely spaced (due to size or physical constraints). The sensor measurements are degraded by additive noise that is not necessarily Gaussian or white.

The linear estimator is applied as a digital filter-bank with a finite impulse-response (FIR) and a variable fractional delay (VFD) of $q_0$ samples. Determination of the time shift for a signal in two spatial channels, is posed as a (linear) minimum mean-squared error (MMSE) problem where the mean squared-error (MSE) is a function of the unknown delay. The function is derived from a sum-of-squared differences (SSD) measure of waveform dissimilarity. Using a VFD filter-bank with a Farrow structure [1], allows the MSE to be expressed as a polynomial in $q_0$. Its stationary points (i.e. local minima and maxima) are then simply determined by finding the roots of a polynomial with a degree of $2K - 3$, where $K$ is the number of sub-filters in the filter bank. This allows the fractional delay that minimises the MSE to be solved directly, without iteratively evaluating candidate delays on an ever-finer grid (then interpolating).

The fact that a Farrow structure leads to an MSE cost-function, that is polynomial in the unknown delay, appears to have been overlooked in the TDE literature; although, the general principle (i.e. the determination of the optimal filter delay that minimises an error or noise power) has previously been used by the author to design smoothers with an infinite impulse response (IIR) in [2] and adaptive beamformers in [3]. An SSD cost-function and Farrow filter-bank is also used to estimate time-delay in [4]; however, derivative-based techniques (e.g. steepest descent and Newton's method) are used to find the delay that minimises the cost. The MSE may also be minimised using a VFD filter in a feedback loop [5],[6],[7],[8]. For instance, a polynomial interpolator with a Farrow structure is used in [9], an all-pass filter with an IIR is used in [10], and a bandpass FIR filter (a windowed Sinc function) is used in [11],[12],[13]. These recursive least-mean-squares (LMS)-style approaches are more appropriate when the time-delay is not constant (e.g. when the sensor axis rotates or then the source is moving); however, careful tuning is required to balance the rate of convergence and the steady-state error variance

The $M$ filter coefficients of the VFD filter are a linear combination of $K$ fixed sub-filters, where the linear factors are powers of the desired passband delay ($q$) for a Farrow structure. In principle, the signal may be shifted arbitrarily (forward or backward) in time; however, noise is amplified, and signals are distorted, as the shift moves away from the centre of the finite time window. Non-central shifts of less than half a sample may be applied with reasonable fidelity.

## 2. TIME-DELAY ESTIMATORS

Consider a system that receives an unknown signal in two adjacent sensors that are identical, and perfectly calibrated. The signal in the 'measurement' channel ($x_1$) is a time-shifted replica of the signal in the 'reference' channel ($x_0$) with measurement noise ($\varepsilon_0$ & $\varepsilon_1$) added to both, i.e.

$x_0(t) = s(t) + \varepsilon_0(t)$ and (1a)

$x_1(t) = s(t - D) + \varepsilon_1(t)$ (1b)

where

$t$ is continuous time (seconds)

$D$ is the time shift (seconds), a constant parameter, with positive and negative shifts corresponding to delays and advances, respectively

$s$ is the signal and

$\varepsilon$ is the internal sensor/receiver noise, which is assumed to be uncorrelated with the signal, spatially uncorrelated, i.e. across the channels, but it may be temporally correlated, i.e. not necessarily Gaussian or white (coloured).

In *passive* direction-finding and location systems (e.g. sonar, radio telescopes/interferometers, radar-warning receivers, gunshot detection/location and tsunami detection/location) the sensor at the spatial origin of the two-element array is chosen to be the reference. In *active* range-finding systems (e.g. sonar, lidar, ultrasound, radar and geo-acoustic sounders) that transmit a signal and receive delayed reflections, the signal in the reference channel is a known 'template', $\varepsilon_0$ is negligible, the signal in the measurement channel is scaled by a propagation-loss factor, and the delays are typically large. The passive case is the primary focus here.

The sensors are (synchronously) sampled at a rate of $F_s$ samples per second (for a sampling period of $T_s = 1/F_s$ seconds per sample) thus in discrete time

$$x_0[n] = s[n] + \varepsilon_0[n] \text{ and} \quad (2a)$$

$$x_1[n] = s[n - q_\Delta] + \varepsilon_1[n] \quad (2b)$$

where

$n$ is the (integer) sample index, with $t = T_s n$ and

$q_\Delta$ is a (fractional) subsample delay, with $D = T_s q_\Delta$.

### 2.1 Generalised Cross-Correlation

TDE problems are commonly solved using generalised cross-correlation (GCC) [14],[15],[16], via the frequency-domain ($\omega$, radians per sample) by applying the Fast Fourier Transform (FFT) and its inverse ($\text{FFT}^{-1}$) using

$$\hat{q}_\Delta = \hat{l} = \underset{l}{\text{argmax}}\, \text{Re}\{\boldsymbol{r}_{01}\} \text{ with} \quad (3a)$$

$$\boldsymbol{r}_{01} = \text{FFT}^{-1}\{\mathcal{X}_0^*(\omega)\mathcal{W}(\omega)\mathcal{X}_1(\omega)\} \quad (3b)$$

where

$\boldsymbol{r}_{01}$ is the $N \times 1$ cross-correlation vector with elements $r_{01}[l]$ for $l = -L \dots + L$

$\mathcal{W}(\omega)$ is a weighting function, designed to minimise signal loss and maximise noise attenuation

$$\mathcal{X}_0(\omega) = \text{FFT}\{x_0[n]\} \text{ and} \quad (3c)$$

$$\mathcal{X}_1(\omega) = \text{FFT}\{x_1[n]\} \quad (3d)$$

($\blacksquare^*$ is the complex-conjugate operator and $\text{Re}\{\blacksquare\}$ returns the real part of the complex argument).

Note that if large delays are expected, the sensor samples should be 'zero-padded' to break the circular boundary condition assumed in the FFT.

For wideband signals and subsample delays, requiring the evaluation of only a few integer lags ($L \cong 1$) the following equivalent time-domain solution, involving summations of filtered and time-shifted cross-channel products, is simpler:

$$r_{01}[l] = \textstyle\sum_{n=L}^{N-L-1} y_0^*[n-l]y_1[n] \text{ for } l = -L \dots + L \text{ or} \quad (4a)$$

$$\boldsymbol{r}_{01} = \boldsymbol{Y}_0^\dagger \boldsymbol{y}_1 \quad (4b)$$

where

$\boldsymbol{y}_1$ is a $N \times 1$ vector of filtered samples from the measurement channel with elements $y_1[n]$

for $n = 0 \dots N-1$ and

$\boldsymbol{Y}_0$ is a $N \times K$ matrix ($K = 2L+1$) with filtered and integer time-shifted samples from the reference channel with elements $Y_0[n,l]$

($\blacksquare^\dagger$ is the Hermitian transpose operator).

The filtering operation

$$\boldsymbol{y}_1 = \boldsymbol{h}_\mathcal{W} * \boldsymbol{x}_1 \quad (5a)$$

is applied in the time domain using

$$y_1[n] = \textstyle\sum_{m=0}^{M-1} h_\mathcal{W}[m]x_1[n-m] \text{ for } n = 0 \dots N-1 \quad (5b)$$

where

$h_\mathcal{W}[m]$ are the coefficients of an FIR filter with impulse response $\boldsymbol{h}_\mathcal{W}$

($*$ is the convolution operator).

The FIR filter has a linear-phase response with a group-delay of $q_\mathcal{W} = (M-1)/2$. It is designed to maximise the signal-to-noise ratio (SNR) in each channel when signal and noise processes are approximately known (e.g. a Wiener filter). For signals with a wide bandwidth, relative to the data-acquisition bandwidth, short impulse responses are sufficient, which further lessens the appeal of a frequency-domain approach. Note that

$$\mathcal{W}(\omega) = \mathcal{H}_\mathcal{W}^*(\omega)\mathcal{H}_\mathcal{W}(\omega) \quad (6a)$$

where

$$\mathcal{H}_\mathcal{W}(\omega) = \text{FFT}\{h_\mathcal{W}[m]\}. \quad (6b)$$

Similarly, for the elements of $\boldsymbol{Y}_0$:

$$y_0[n] = \textstyle\sum_{m=0}^{M-1} h_\mathcal{W}[m]x_0[n-m] \text{ then} \quad (7a)$$

$$Y_0[n,l] = y_0[n-l]. \quad (7b)$$

Note that the start of each sample buffer or data vector (i.e. $\boldsymbol{x}_0$ & $\boldsymbol{x}_1$) should be 'data-padded' with $M-1$ additional samples to skip the filter start-up transient. For subsample delays, using $L = 1$, followed by three-point quadratic interpolation around $r_{01}[\hat{l}]$, may be sufficient.

### 2.2 A Novel MMSE Delay Estimator

The subsample GCC estimator described above may be improved upon using an MSE that is an explicit function of the unknown (fractional) delay using

$$\hat{q}_0 = \underset{q_0}{\text{argmin}}\, P_\varepsilon(q_0) \text{ then } \hat{q}_\Delta = \hat{q}_0 - q_1 \quad (8)$$

where

$P_\varepsilon(q_0)$ is the MSE 'cost' function – a measure of signal misalignment across both channels

$\hat{q}_0$ is an estimate of the reference delay that aligns the signal in both channels

$q_0$ is a variable delay applied (via a filter-bank) to the samples from the reference channel and

$q_1$ is a fixed delay applied (via a linear-phase filter) to the samples from the measurement channel.

The proposed MSE functional is a sum-of-squared differences (SSD) dissimilarity metric, evaluated via a VFD filter, realised using an FIR filter-bank with a Farrow structure. For real or complex waveforms, the error function is evaluated as follows:

$$P_\varepsilon(q_0) = \tfrac{1}{N}\boldsymbol{\varepsilon}^\dagger\boldsymbol{\varepsilon} \text{ with} \tag{9a}$$

$$\boldsymbol{\varepsilon} = \boldsymbol{y}_1 - \boldsymbol{Y}_0^{\mathrm{VFD}}\boldsymbol{b}(q_0) \tag{9b}$$

where

$$\boldsymbol{y}_1 = \boldsymbol{h}_1 * \boldsymbol{x}_1 \tag{9c}$$

$\boldsymbol{Y}_0^{\mathrm{VFD}}$ is an $N \times K$ matrix with columns $\boldsymbol{h}_k^{\mathrm{VFD}} * \boldsymbol{x}_0$

for $k = 0 \ldots K-1$ and

$\boldsymbol{b}(q_0)$ is a $K \times 1$ vector with elements

$b[k] = q_0^k$.

Substitution of (9b) into (9a), with the factor of $1/N$ and the VFD superscript omitted, yields

$$P_\varepsilon(q_0) = \tag{10a}$$

$$\boldsymbol{y}_1^\dagger\boldsymbol{y}_1 - \boldsymbol{y}_1^\dagger\boldsymbol{Y}_0\boldsymbol{b}(q_0) - \boldsymbol{b}^\dagger(q_0)\boldsymbol{Y}_0^\dagger\boldsymbol{y}_1 + \boldsymbol{b}^\dagger(q_0)\boldsymbol{Y}_0^\dagger\boldsymbol{Y}_0\boldsymbol{b}(q_0)$$

which has the polynomial form

$$P_\varepsilon(q_0) = \textstyle\sum_{k=0}^{2K-2} c_k q_0^k. \tag{10b}$$

Differentiating $P_\varepsilon(q_0)$ with respect to $q_0$ yields the polynomial

$$P_\varepsilon'(q_0) = \textstyle\sum_{k=0}^{2K-3} \bar{c}_k q_0^k. \tag{11a}$$

Stationary points of $P_\varepsilon(q_0)$, i.e. local minima (or maxima) are found by solving

$$P_\varepsilon'(q_0) = 0. \tag{11b}$$

Following the problem statement in (8) for subsample TDE estimation, $\hat{q}_0$ is set equal to the real root of $P_\varepsilon'(q_0)$ that is closest to $q_1$. The TDE estimate is then converted from samples to seconds using

$$\hat{D} = T_s\hat{q}_\Delta. \tag{12}$$

For this subsample delay estimator, the $k$th column of $\boldsymbol{Y}_0^{\mathrm{VFD}}$ is now formed by convolving $\boldsymbol{x}_0$ with the $k$th sub-filter $\boldsymbol{h}_k^{\mathrm{VFD}}$ of a VFD filter with a Farrow structure. Farrow showed that the impulse response of a low-pass FIR VFD filter $\boldsymbol{h}^{\mathrm{VFD}}$ (with a length of $M$ samples and a non-linear phase response) may be realised using a set of fixed linear-phase sub-filters $\boldsymbol{H}^{\mathrm{VFD}}$ and a vector $\boldsymbol{b}$ that is a known polynomial function of the desired passband delay $q$, as defined in (9), using

$$\boldsymbol{h}^{\mathrm{VFD}} = \boldsymbol{H}^{\mathrm{VFD}}\boldsymbol{b}(q) \tag{13}$$

where

$\boldsymbol{H}^{\mathrm{VFD}}$ is a $M \times K$ matrix with the sub-filters $\boldsymbol{h}_k^{\mathrm{VFD}}$ as its columns.

The coefficients of $\boldsymbol{h}_1$ that are used to form $\boldsymbol{y}_1$ in (9c) are derived using the same procedure that is used to form $\boldsymbol{Y}_0^{\mathrm{VFD}}$; however, a fixed delay of $q_1 = (M-1)/2$ is used for a linear-phase low-pass FIR filter. Note that $q_\Delta$ is estimated by determining the delay applied to the reference channel that matches the unknown propagation delay/advance applied to the measurement channel. A fixed processing delay ($q_1$) is applied to the measurement channel to ensure that the (causal) VFD FIR filter used to process the reference channel has an approximately linear phase-response over the signal band and a reasonably flat magnitude-response over the signal and noise bands (of unity and zero, respectively).

## 3. VARIABLE FRACTIONAL-DELAY FILTERS

In a 1988 conference paper, Farrow described a simple FIR filter-bank with a low-pass or all-pass magnitude response and a group delay ($q$) that may be shifted forward or backward by scaling the sub-filters using a coefficient vector that is a polynomial in the desired delay [1]. His paper focused on the simplicity and elegance of this then un-known signal-processing structure. The sub-filter coefficients were found by minimising the integral of the squared error over the full range of fractional delays ($q_\Delta = \pm 0.5$, samples) and from dc up to the "upper useful frequency" ($f_c = 0.33$, cycles per sample). The response of the filter for various subsample shifts is shown in Figure 1.

The Farrow structure is now ubiquitous in multi-rate signal-processing systems. A variety of procedures for the derivation of sub-filter coefficients have since been described [17],[18],[19]. Time-domain polynomial interpolation (e.g. Lagrange) is an obvious and intuitive approach; however, frequency-domain techniques are favoured in the recent literature and are arguably a more natural way of measuring and modelling generic wideband and narrowband processes (e.g. signal & noise) and systems (e.g. sensors & samplers). For instance: minimisation of the $l_2$ error norm (i.e. least-squares or integral squared-error methods) [20]; $l_\infty$ error norm (i.e. minimax methods for an equi-ripple response) [21],[22]; or the more general $l_p$ error norm [23]. As derivatives around $\omega = 0$ are monomials around $m = 0$ [2],[3],[24], dc derivative constraints and a (white or coloured) noise model are a simple way of specifying and deriving suitable sub-filter coefficients. Any of these filter-bank designs may be used to evaluate the polynomial MSE cost function ($P_\varepsilon$) considered in the previous subsection. However, for simplicity and brevity, Farrow's original VFD filters are re-used here to process the simulated data in the next section.

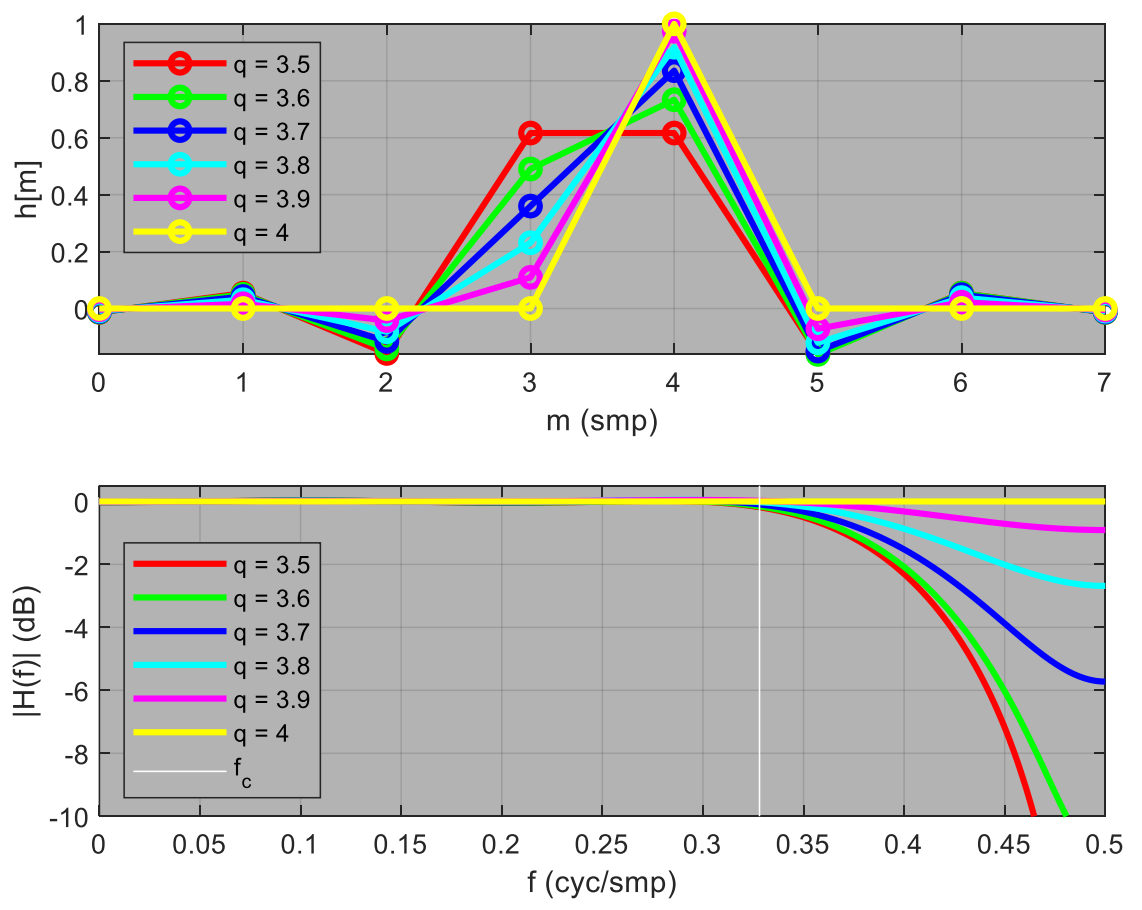

**Figure 1.** Impulse response (top) and magnitude response (bottom) of Farrow's FIR VFD for various fractional group delays ($q$).

## 4. SIMULATIONS

The Monte-Carlo simulations presented here are used to illustrate and compare the TDE estimation techniques described above: Generalised Cross Correlation (GCC) and MMSE estimation using a filter bank with a Farrow structure (VFD). Time shifts ($q_\Delta$) were randomly generated using a uniform distribution over the $[-1/2, +1/2]$ interval and all estimates ($\hat{q}_\Delta$) were limited, so they did not exceed $\pm 1/2$.

The (real) signal was generated by passing a randomly generated sequence of Gaussian noise through a low-pass filter designed using a Gaussian-windowed Sinc function. The FIR filter had a length of 25 samples, a Gaussian standard-deviation of 4 samples, and a Sinc cut-off frequency of $f_c = \pm 0.33$ (cycles per sample). Subsample shifts were applied to the signal in the measurement channel by shifting the time argument of the Gaussian and the Sinc functions. White Gaussian noise was added to both channels, and four signal-to-noise ratio (SNR) scenarios were considered: 30, 20, 10 & 0 dB. One thousand (real-valued) samples in each channel were generated and processed (see Figure 2). The root mean-squared error (RMSE) metric was computed over ten-thousand random instantiations of each scenario for both techniques (see Table I).

**Table I. RMSE (samples) of Time-Delay Estimators as a Function of SNR (dB)**

| | **30** | **20** | **10** | **0** |
|---|---|---|---|---|
| **GCC** | 0.030 | 0.030 | 0.033 | 0.059 |
| **VFD** | 0.001 | 0.004 | 0.020 | 0.142 |

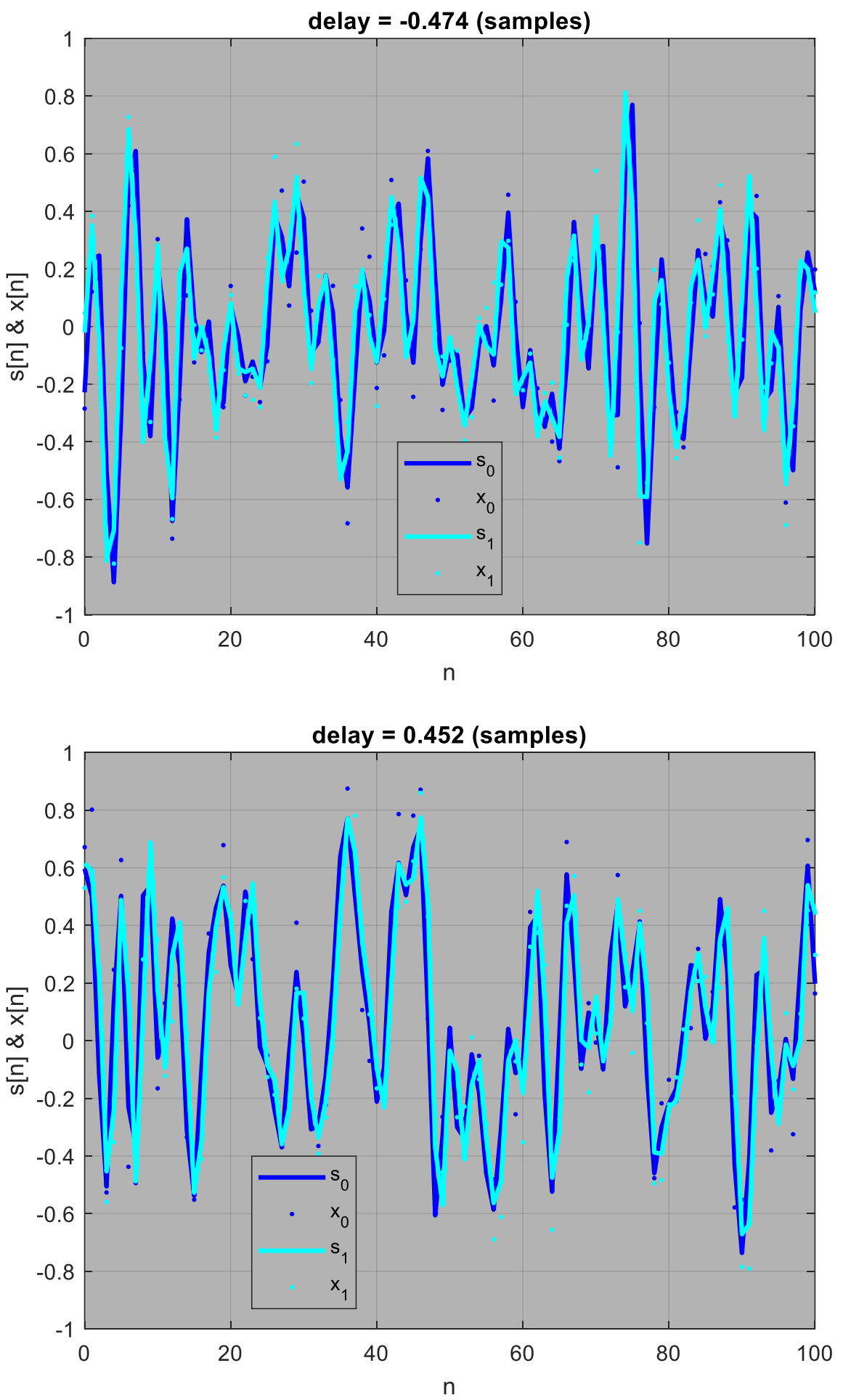

**Figure 2.** Two random instantiations of the 10 dB scenario.

## DISCUSSION AND CONCLUSION

The simulations indicate that the proposed time-delay estimator behaves as intended. The VFD and GCC filter banks (both with $M = 8$) apply the same (band limited) signal in (white) noise model; however, the VFD approach (with $K = 4$) leverages it for greater accuracy in relatively clean conditions (30 to 10 dB). GCC (with $K = 3$) further assumes that the signal cross-correlation ($\boldsymbol{r}_{01}$) is approximately parabolic near zero, for greater robustness at low SNR (0 dB).